\documentclass[twocolumn,aps,prl,showpacs,superscriptaddress]{revtex4}
\usepackage{graphicx}
\usepackage{amssymb} 
\usepackage{url}     
\usepackage{bm}      
\usepackage{color}

\newcommand{\bise}{Bi$_2$Se$_3$}

\newcommand{\vtg}{$V_{TG}$}

\newcommand{\didv}{$dI/dV$}
\newcommand{\vsd}{$V_{SD}$}
\newcommand{\vti}{$V_{TI}$}

\newcommand{\vbg}{$V_{BG}$}
\newcommand{\sio}{SiO$_2$}

\begin{document}

\author{H. Steinberg}
\affiliation{Department of Physics, Massachusetts Institute of Technology, Cambridge, MA 02139, USA}
\affiliation{Racah Institute of Physics, The Hebrew University, Jerusalem 91904, Israel}
\author{L. A. Orona}
\affiliation{Department of Physics, Massachusetts Institute of Technology, Cambridge, MA 02139, USA}
\author{V. Fatemi}
\affiliation{Department of Physics, Massachusetts Institute of Technology, Cambridge, MA 02139, USA}
\author{J.D. Sanchez-Yamagishi}
\affiliation{Department of Physics, Massachusetts Institute of Technology, Cambridge, MA 02139, USA}
\author{K. Watanabe}
\affiliation{Advanced Materials Laboratory, National Institute of Material Science, 1-1 Namiki, Tsukuba 305-0044, Japan}
\author{T. Taniguchi}
\affiliation{Advanced Materials Laboratory, National Institute of Material Science, 1-1 Namiki, Tsukuba 305-0044, Japan}
\author{P. Jarillo-Herrero}
\affiliation{Department of Physics, Massachusetts Institute of Technology, Cambridge, MA 02139, USA}


\title{Tunneling in Graphene-Topological Insulator Hybrid Devices}

\begin{abstract}
Hybrid graphene-topological insulator (TI) devices were fabricated using a mechanical transfer method and studied via electronic transport. Devices consisting of bilayer graphene (BLG) under the TI \bise\ exhibit differential conductance characteristics which appear to be dominated by tunneling, roughly reproducing the \bise\ density of states. Similar results were obtained for BLG on top of \bise, with 10-fold greater conductance consistent with a larger contact area due to better surface conformity. The devices further show evidence of inelastic phonon-assisted tunneling processes involving both \bise\ and graphene phonons. These processes favor phonons which compensate for momentum mismatch between the TI $\Gamma$ and graphene $K, K'$ points. Finally, the utility of these tunnel junctions is demonstrated on a density-tunable BLG device, where the charge-neutrality point is traced along the energy-density trajectory. This trajectory is used as a measure of the ground-state density of states.
\end{abstract}

\maketitle



\date{\today}

In topological insulators (TIs), the protected surface state is bound to the interface between materials with different bulk topological invariants \cite{Hasan_Kane_2010}. Although the surface state will remain gapless as long as the symmetries protecting the bulk topological invariant are retained, its dispersion and location depend on the band-structures of the two interfacing materials. A model system allowing the investigation of such interface properties is the graphene-TI hybrid, where graphene resides in immediate proximity to the TI, separating it from the vacuum or from another trivial dielectric. Indeed, as part of a 3D continuum, graphene is topologically trivial. However, with a significant spin-orbit (SO) term, graphene could become a 2D TI \cite{Kane_Mele_2005}.

Recent studies \cite{Jin_2013,Liu_2013, Kou_2013, Zhang_Rossi_2014} which theoretically investigate the properties of such hybrids, generally assume that the graphene and TI bands couple strongly, leading to fundamental modifications to the graphene band structure, which is expected to inherit an enhanced SO coupling \cite{Jin_2013} and attain non-trivial spin-textures \cite{Zhang_Rossi_2014}. When sandwiched between two ultra thin layers of \bise, the graphene layer may become a 2D TI \cite{Kou_2013}. However, as the surface states of TIs such as \bise\ are centered at the $\Gamma$ point in $k$-space, they should be prevented by momentum conservation from hybridizing with the graphene bands, centered at the $K, K'$ points. This difficulty is often resolved theoretically by assuming that periodicity at the interface would cause the graphene Brillouine zone (BZ) to fold \cite{Jin_2013,Liu_2013, Kou_2013, Zhang_Rossi_2014}, bringing significant spectral weight to the $\Gamma$ point.


Here we present an experimental study of graphene-TI hybrid devices fabricated using a mechanical transfer technique \cite{Dean_hBN_2010} which allows the fabrication of vertical heterostructures involving a wide variety of  van-der-Waals materials. In our devices, which consist of either monolayer (MLG) or bilayer graphene (BLG) stacked above or below the TI \bise, the graphene and TI layers are weakly coupled, with transport between them governed by tunneling. The devices reveal a rich inelastic spectrum consisting of graphene and \bise\ phonons. Finally, we demonstrate the utility of the interface tunnel-junction to probe the density-dependent evolution of spectral features in BLG. 


\begin{figure}
\label{Figure_1}
\begin{center}
\includegraphics[width = 80mm]{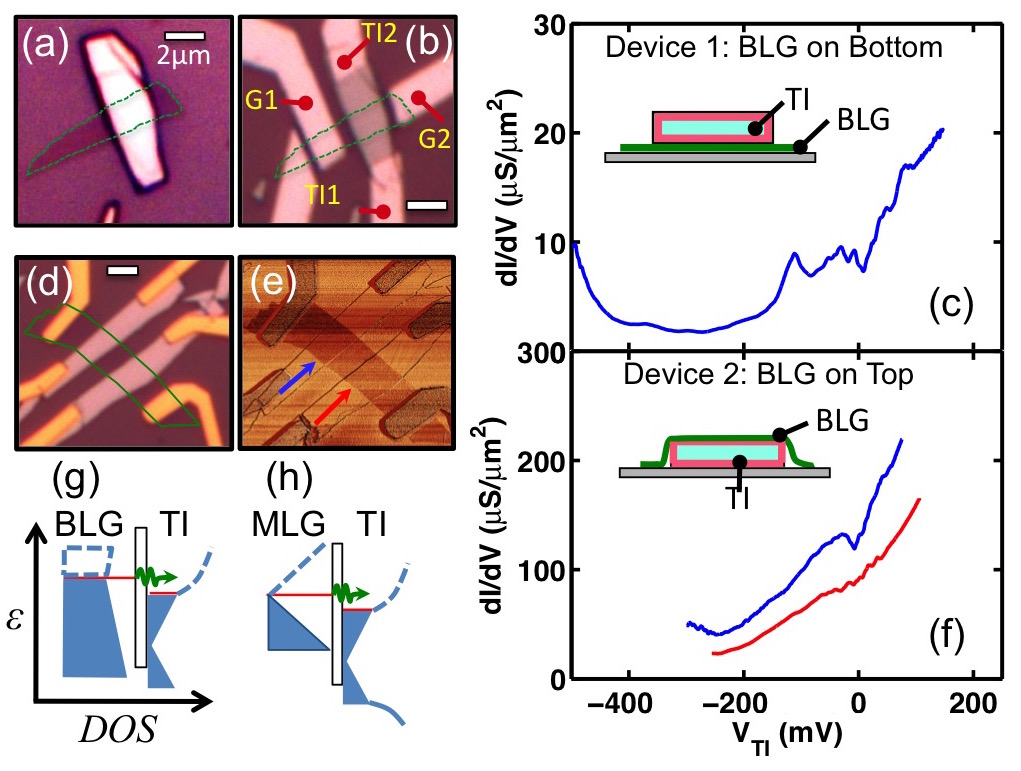}
\vspace{0cm}
\caption{(a) Optical image of Device 1, consisting of BLG on the bottom and \bise\ on top. BLG outline marked in green. (b) Device 1 with contacts. Bar is $2\mu m$ long. (c) $dI/dV$ vs. $V_{TI}$ of Device 1 (source on the \bise.). Inset: Schematic showing the BLG (green) underneath the TI (teal, outline representing the TI surface states). (d) Device 2: Single BLG flake deposited on top of two \bise\ flakes (e) AFM phase image of Device 2, arrows mark the two separate junctions (f) $dI/dV$ vs. $V_{SD}$ of the two junctions in Device 2. Top junction: blue, bottom: red. Inset: Schematic showing graphene or BLG on top of the TI. (g,h) Annotations illustrating the DOS alignment in tunneling between BLG (g) and MLG (h) and a TI. All $dI/dV$ data are normalized per $\mu m^2$.
}
\end{center}
\vspace{-0.8 cm}
\end{figure}


Fig. 1(a) shows Device 1, where a \bise\ flake is transferred on top of a BLG flake (outlined in green) deposited on \sio. The flakes are independently contacted by evaporated metallic electrodes (Fig. 1(b)). The BLG-TI junctions are studied by measuring the differential conductance \didv\ vs. applied bias voltage \vsd\ with the source on one material and the drain on the other (TI1 to G1). A four-probe geometry is realized by measuring the voltage across the opposing contacts ($V_{TI}$ = V(TI2)-V(G2)). The \didv\ trace at $T = 4K$ of Device 1 (Fig. 1(c)) shows a pronounced suppression at negative $V_{TI}$ values, similar to STM measurements taken on \bise\ \cite{Alpichshev_2012, Cheng_LL_MBE_2010, Hanaguri_2010}, where \didv\ is proportional to the density of states (DOS) and the suppression at negative bias is associated with the bulk gap of the \bise. The BLG-TI interface therefore behaves as an effective  tunnel junction although no intentional barrier was placed between the materials. It is a remarkably robust  junction, maintaining a stable signal while sustaining high voltage biases exceeding 0.5 V at negative bias.

The tunneling functionality of the graphene-\bise\ interface could be a consequence of \bise\ oxidation \cite{Kong_Oxidation_2011}. In over 20 devices studied, the interface resistance varied from 10 M$\Omega\mu m^2$ to 10 $k\Omega\mu m^2$, which could be associated with varying degrees of oxidation. The stability and high bias achieved by graphene-TI junctions, however, do not favor this explanation -- oxide-based junctions rarely function at biases in excess of 200 meV \cite{Tsui_1970}. Alternatively, the variation in interface resistance could arise due to differences in the effective contact area: graphene conforms to the underlying substrate and has angstrom-scale height variations when deposited on \sio\, resulting in an effective partial contact area. To test this we studied devices where BLG is deposited \textit{on top} of \bise: in Device 2 a single BLG flake covers two \bise\ flakes (Fig. 1(d-e)). Measured separately (red and blue curves in panel (f)), the two junctions exhibit comparable \didv\ characteristics which are very similar to the ``graphene on bottom" devices (albeit with smaller bias range).

The interface conductivity of ``graphene-on-top" junctions is an order of magnitude greater than ``graphene on bottom" ones, indicating that the effective contact area is a plausible source of variation. It is not clear, however, what is the origin of the generic tunneling functionality. An interesting possibility is that the tunnel barrier is associated with the lack of direct chemical bonding between the two layers. Inter-layer tunnel barriers in layered materials is observed in high Tc superconductors \cite{Kleiner_1994} and in vdW materials such as 4Hb-$TaS_2$ \cite{Wattamaniuk_1975}. Incoherent interlayer transport was also reported in stacked twisted bilayer graphene devices \cite{Kim_PRL_2013}, and graphene layers could be weakly coupled to underlying graphite \cite{Li_Andrei_2007}. Testing this hypothesis would require further investigation.

\begin{figure}
\label{Figure_2}
\begin{center}
\includegraphics[width = 80mm]{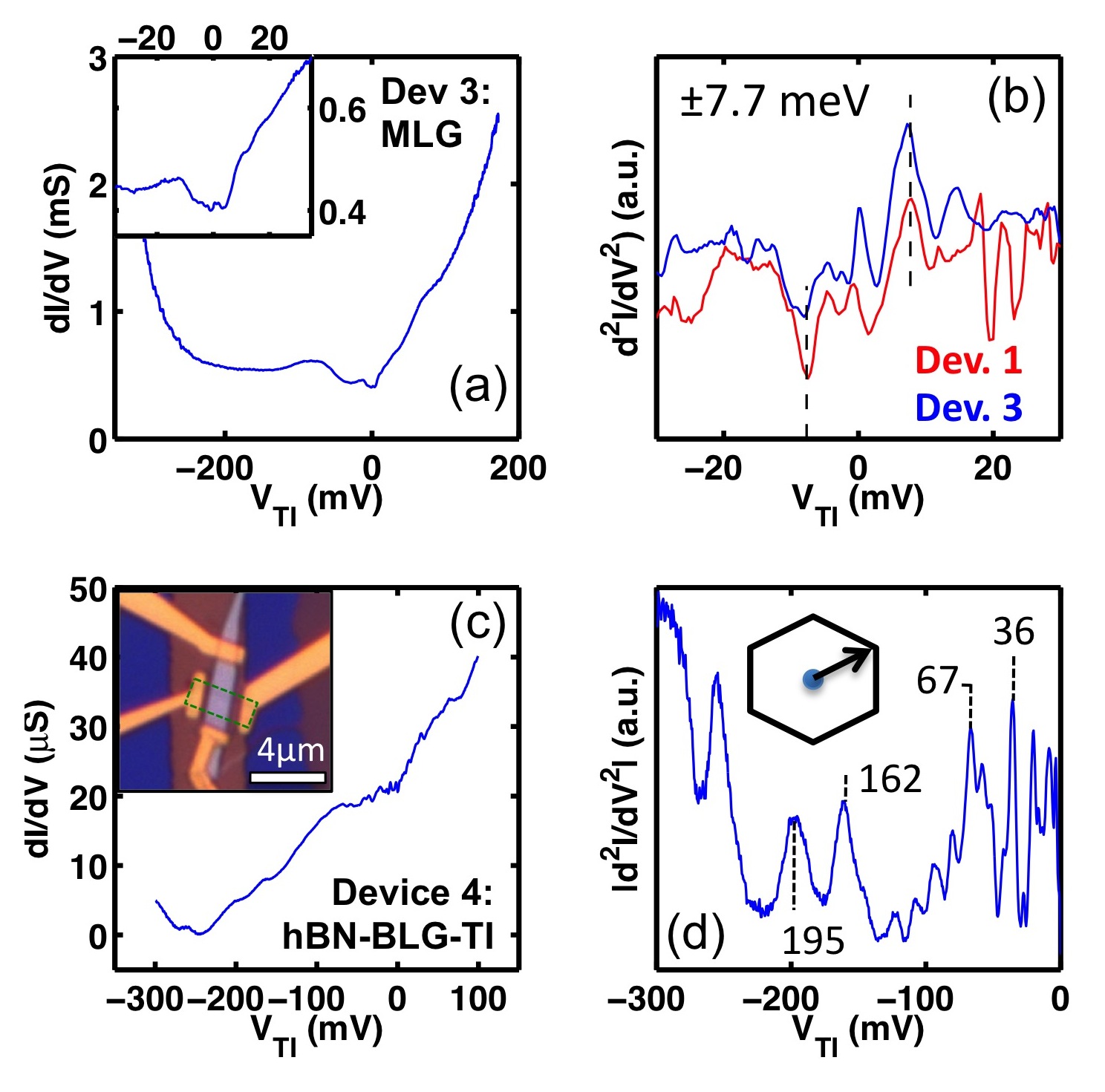}
\vspace{0cm}
\caption{(a) $dI/dV$ vs. $V_{TI}$ of Device 3:  \bise\ on top of monolayer graphene. Inset: Closeup of the data showing the zero-bias suppression associated with inelastic tunneling features. (b) $d^2I/dV^2$ vs. $V_{TI}$ of Device 1 (BLG) and Device 2 (MLG). Both devices exhibit a peak in the 2nd derivative, corresponding to an inelastic excitation, at 7.7 meV, associated with a \bise\ optical phonon (c) $dI/dV$ vs. $V_{TI}$ of Device 4, which is a vertical heterostructure of a hBN substrate, BLG, and \bise\ on top. Inset: Image of the device (Bar is $4 \mu m$). (d) $|d^2I/dV^2|$ vs. $V_{TI}$ of Device 4, exhibiting a rich spectrum of inelastic features. The energies marked in the figure are discussed in the text. Inset: Phonon assisted tunneling in momentum mismatched materials: Tunneling from the TI states at the $\Gamma$ point to graphene states at the $K, K'$ points is assisted by BZ-edge phonons. }

\end{center}
\vspace{-0.8 cm}
\end{figure}

As expected, the \didv\ characteristics vary with the choice of MLG vs. BLG. The tunneling current is expressed as a convolution of the graphene DOS $\rho_{G}(\varepsilon)$ and the TI DOS $\rho_{TI}(\varepsilon)$ \cite{wolf_book}
\begin{equation}\label{Tun}
I \sim A\int_0^{eV}\rho_{G}(\varepsilon-eV)\rho_{TI}(\varepsilon)|t(\varepsilon)|^2d\varepsilon
\end{equation}
where $A$ is the effective overlap area and $t(\varepsilon)$ stands for the tunnel coupling through the barrier. This expression is qualitatively depicted in the schematics in Fig. 1(g,h), showing $\rho_G$ on the left and $\rho_{TI}$ (including bulk and surface) on the right. For BLG devices (panel (g)) $\rho_{G}(\varepsilon)$ is relatively featureless and can be factored out of the integral, so the \didv\ curve effectively probes $\rho_{TI}(\varepsilon)$. This explains why BLG-TI \didv\ traces are similar to STM measurements of \bise, and suggests that BLG may be useful as a tunneling electrode for probing other vdW systems. MLG-TI devices (panel (h)) have energy-dependent $\rho(\varepsilon)$ on both sides of the barrier, and Device 3 (Fig 2(a)) indeed exhibits a very different \didv\ characteristic, with a stronger suppression around zero-bias which we associate with the graphene Dirac point.

\begin{figure*}
\label{Figure_3}
\begin{center}
\includegraphics[width = 176mm]{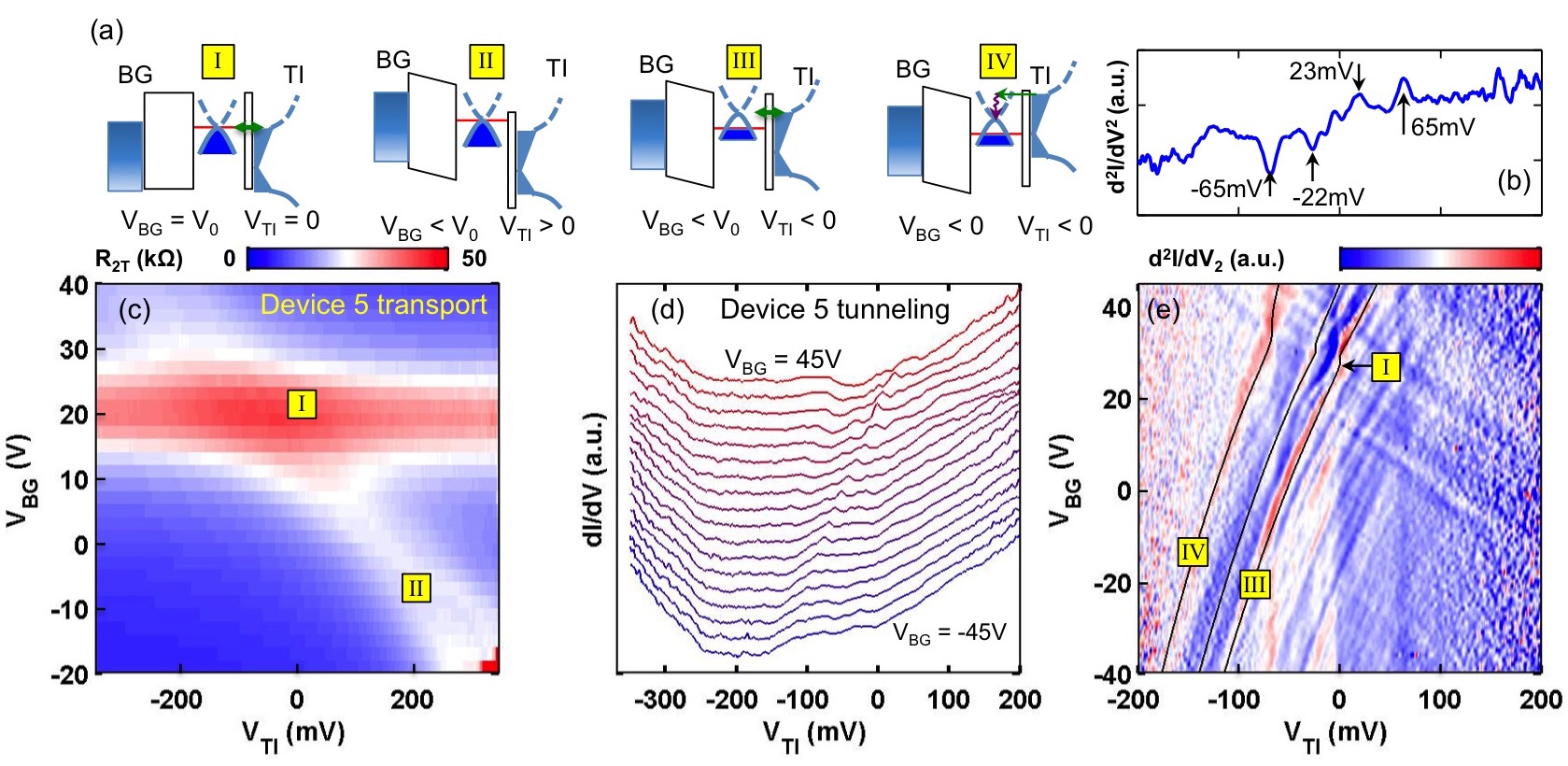}
\vspace{0cm}
\caption{(a) Schematics of gate-dependent tunneling and transport processes. (I) BLG at $n = 0$, \vti\ at zero bias. (II) tracing the CNP in a transport measurement: \vbg\ and \vtg\ are balanced to keep $n=0$. (III) tunneling into the CNP: $V_{BG}<0$ tunes the BLG to $n>0$, $V_{TI} < 0$ keeps tunneling to the CNP (IV) phonon-assisted tunneling into the CNP. (b) $d^2I/dV^2$ vs. \vti\ obtained from integrating data in (e). (c) In-plane electronic transport vs. \vbg\ and \vti. Annotation II marks the CNP diagonal trajectory. Annotation I marks the point where this trajectory crosses \vti\ = 0. (d) $dI/dV$ vs. \vti\ at a range of gate voltage (5V interval). (e) $d^2I/dV^2$ vs. \vti\ and \vbg. Annotations mark the elastic process (III) and phonon-assisted processes (IV). The lines superimposed on the data are fits to the trajectories of elastic and inelastic tunneling into the CNP \cite{SI}.}
\end{center}
\vspace{-0.5cm}
\end{figure*}

Tunnel junctions are useful in measuring inelastic spectra, where the onset of processes such as phonon-assisted tunneling appears as step-increase features in the differential conductance. These are detectable as peaks in the 2nd derivative $d^2I/dV^2$ at bias voltages corresponding to the phonon energies. It is possible to differentiate between \bise\ and graphene phonons because their respective spectra span different energy ranges (up to 20 meV for \bise\ \cite{Zhu_2011}, and up to 200 meV for graphene \cite{Wirtz_Rubio_2004}).
Several devices exhibit well-developed inelastic spectra: Device 3 (Fig. 2(a)), has sharp step-like features in \didv\  close to zero (inset). The corresponding $d^2I/dV^2$ plot (Fig. 2(b)), shows peaks at $\pm$7.7 meV which coincide in Device 3 (MLG) and in Device 1 (BLG). We associate this feature with a \bise\ surface optical phonon previously identified using Helium scattering at the same energy \cite{Zhu_Phonons_2012}. This phonon plays a role in suppressing surface transport \cite{Costache_2014}, and inelastic tunneling data may be useful in probing its coupling to surface electrons.


Signatures of graphene phonons, appearing at higher bias voltages, are found in numerous devices. For example, in Device 4 (Fig. 2(c,d)), a higher quality device fabricated by deposition of BLG on h-BN followed by transferring \bise\ on top. The $d^2I/dV^2$ plot shows a prominent peak at 67 meV, corresponding to the energy of the graphene ZA/ZO mode at the $K, K'$ points. This phonon is seen in many graphene and graphite tunneling experiments both in devices \cite{Amet_DGG_2012} and in STM \cite{Zhang_2008, Brar_2010}, and is generally believed to assist the tunneling process by providing the momentum required to inject an electron to the $K, K'$ points. In our devices, where graphene and \bise\ Fermi surface momenta are highly mismatched, it is likely that the same phonon is required to allow momentum-conserving tunneling from low energy \bise\ states, centered at the $\Gamma$ point, to the graphene $K, K'$ points (Schematic in inset to Fig. 2(d)). Device 4 also exhibits a feature at 162 meV associated with the LA/LO mode, also a $K, K'$ phonon, and a feature at 195 meV associated with the TO mode at $\Gamma$. Other features (e.g. a phonon at 36 meV) could be associated with the h-BN substrate \cite{Amet_DGG_2012}.

In the rest of this Letter we focus on Device 5, which has the same geometry as Device 1: TI on top of a density-tunable BLG flake. Device 5 can be measured by graphene in-plane transport (G1 to G2) or by tunneling (T1 to G1). In-plane resistance $R_{2T}(V_{TI},V_{BG})$ is presented in Fig. 3(c), showing a resistance peak when the chemical potential crosses the BLG CNP. The \bise\ electrode acts as a well-behaved gate, and the small tunneling current does not interfere with the in-plane measurement. The resistance map is typical of doubly-gated graphene \cite{Huard2007,Thiti_PRL_2010}, where the diagonal feature corresponds to a high resistance state in the BLG region underneath the TI. This feature intercepts $V_{TI}=0$ at $V_{BG} = V_0 = 20 V$, indicating that the BLG is p-doped and the CNP is energetically mismatched from the TI DP (Fig. 3(a) annotation I). Its slope is dictated by the requirement that the top and bottom gates compensate each other's charge (annotation II) and therefore follows the ratio $C_{BG}/C_{TI}$ ($C_{TI}$ and $C_{BG}$ are the graphene-TI and graphene-back gate capacitances, respectively). Using this relation, we extract the geometric capacitance of the TI-BLG junction $C_{TI} = 1.3\times 10^{-2}F/m^2$, 110 times greater than the back-gate capacitance. Both the relative permittivity of the interlayer medium, $\varepsilon$, and the graphene-\bise\ effective distance, $d$, are unknown. However, the measured value of $C_{TI}$ fixes their ratio to $1.5$ nm$^{-1}$. As the interface is a tunnel junction, $d$ is unlikely to exceed $3$ nm, setting a limit of $\varepsilon < 5$.

Gate-dependent tunneling measurements are presented in Fig. 3(d-e): Such measurements are sensitive to energy shifts in the spectral features due to the changes in density. In Fig. 3(d) it is clear that the underlying structure of the \bise\ gap at negative \vti\ is retained while a set of features closer to zero bias evolve with \vbg. To trace these features we plot the second derivative $d^2I/dV^2$ as a color map in Fig 3(e). The data exhibit a set of diagonal features with opposite slopes and gate-independent features which appear as faint vertical lines. The latter gate-independent features are enhanced by averaging over all back-gate values (Fig. 3(b)), where we again find the ZA/ZO phonon features at $\pm$65 meV. The 22 meV feature coincides with the energy of the \bise\ $A^2_{1g}$ phonon \cite{Zhu_2011}.

\vbg\ changes the graphene density, vertically shifting the band-structure (Fig. 3(a), annotation III). To trace a spectral feature such as the CNP, a voltage $\delta V_{TI}$ has to be applied to compensate for the density-induced change in chemical potential, $\delta\mu_G$ . This is formulated as:
\begin{equation}\label{tunneling_CNP_condition}
e\delta V_{TI} = \delta \mu_G
\end{equation}

To trace spectral features on the \vti-\vbg\ plane, we have to note that planar tunneling electrodes are large capacitors which charge the graphene layer at finite bias, as discussed earlier. The equations governing the charging of the graphene layer connect the incremental gate voltages $\delta V_{TI,BG}$ to the incremental charges on the back-gate and TI $\delta n_{BG,TI}$:
\begin{equation}\label{eq_charging}
-e\delta n_{i} = C_{i}\left(\delta V_{i}-\frac{\delta\mu_G}{e}\right)
\end{equation}
where $i=BG,TI$. 

Interestingly the CNP tunneling condition in Eq. \ref{tunneling_CNP_condition} also ensures that there will be no extra charge accumulated on the TI ($\delta n_{TI} = 0$), as the argument in parenthesis in Eq. \ref{eq_charging} remains zero. More generally, this means that constant energy features lie on trajectories which keep the charge on the tunneling electrode fixed. In the Supplementary Information \cite{SI} we derive the CNP-tunneling trajectory and find its slope to be:
\begin{equation}\label{eq_trajectory}
\frac{\delta V_{TI} }{\delta V_{BG}} = \frac{C_{BG}}{C_Q+C_{BG}}
\end{equation}
where $C_Q = e^2\rho$ is the quantum capacitance of the BLG. The slope is independent of $C_{TI}$, consistent with the induced charge on the TI remaining zero.

As a result, Eq. \ref{eq_trajectory} can be used to evaluate $C_Q$ by fitting the slope of the features appearing in the figure. The most prominent feature in Fig. 3(e), marked by ``III", is likely the charge neutrality point, which attains a cusp in the DOS due to a finite displacement field (annotation I). The slope of this feature is not constant, and appears to vary along the trajectory (Fig. S1). This indicates that $C_Q$, and hence the DOS, changes with density, as expected for the hyperbolic dispersion of BLG \cite{DasSarmaReview_2011, CastroNeto_RMP_2009, McCann_BLG_2006}. 

We fit the expected trajectory (Fig. 3(e), feature ``III"). To calculate this fit it is crucial to acknowledge that the BLG dispersion varies within the $V_{BG}-V_{TI}$ plane, where at each point the displacement field is different. The calculation, described in the SI \cite{SI}, is carried out self-consistently, using an approximate trajectory for calculating the displacement field at each value of \vbg\ and \vti. We then extract $n(\varepsilon)$, $\rho(n)$ and finally $C_Q$ and integrate the trajectory from the slope in Eq. \ref{eq_trajectory}. The calculation is refined iteratively. Using the Fermi velocity $v_F$ as a fitting parameter, the full trajectory can be reproduced with $v_F = 1.06 \times 10^6 m/s$, in good agreement with non-interacting values for BLG.




Multiple replica (marked by IV in Fig. 3(e)) of the CNP tunneling feature, appear as lines running parallel to the elastic feature (``IV"). They are associated with phonon-assisted inelastic tunneling processes to the CNP. Unlike the gate-independent phonon-onset features discussed above, where tunneling takes place to the Fermi energy, these features represent phonon-assisted tunneling to some sharp spectral feature (here the CNP), and evolve with the gate in parallel to the elastic tunneling feature. Their trajectories, presented in Fig. S1, should depend on the elastic feature, after accounting for modified displacement fields at higher bias. However, to fit the actual data we find that the Fermi velocities have to be modified in each of the inelastic features (assuming interlayer coupling $t_\perp$ remains fixed). The modified Fermi velocities are $0.98\times 10^6$ m/s for the -22 mV feature, and $0.9\times 10^6$ m/s for the -65 mV feature. This points, perhaps, to velocity renormalization.


In summary, the graphene-\bise\ interface is a high quality tunnel junction which can be integrated in to a density-tunable device. Further studies are required to address the effect of junction properties such as crystallographic orientation, interface quality and the effect of \bise\ oxidation. Phonon-assisted tunneling is observed at finite bias, with specific phonon-activation processes which bridge the mismatch in crystal-momentum between the two materials. Nevertheless, tunneling in general is not momentum conserving, probably due to junction inhomogeneity or other scattering processes. The role of the surface state in the tunneling process is also an open question: where some device (Device 1) seem to reveal evidence of bulk states in the tunneling signal, others (Device 4) appear to be mostly surface-dominated.

This work was supported by the DOE, Basic Energy Sciences Office, Division of Materials Sciences and Engineering, under Award No. DE-SC0006418. This work made use of the Materials Research Science and Engineering Center Shared Experimental Facilities supported by NSF under Grant No. DMR-0819762. Sample fabrication was performed partly at the Harvard Center for Nanoscale Science supported by the NSF under Grant No. ECS-0335765.

\bibliographystyle{aps}

\end{document}